\documentclass[a4paper,11pt]{article}
\usepackage{pos}

\usepackage{blindtext}
\usepackage{microtype}

\title{The Spectrum of Global Axion Strings}

\author*[a]{Mathieu Kaltschmidt}
\author[a,b]{Javier Redondo}
\author[c]{Ken'ichi Saikawa}
\author[a]{Alejandro Vaquero}

\affiliation[a]{CAPA \& Departamento de Física Teórica, Universidad de Zaragoza,\\
 Pedro Cerbuna 12, 50009 Zaragoza, Spain}
 \affiliation[b]{Max-Planck-Institut für Physik (Werner-Heisenberg-Institut),\\
Föhringer Ring 6, 80805 München, Germany}
\affiliation[c]{Institute for Theoretical Physics, 
			Kanazawa University,\\
			Kakuma-machi, Kanazawa, 
			Ishikawa 920-1192, 
			Japan}

\emailAdd{mkaltschmidt@unizar.es}
\emailAdd{jredondo@unizar.es}
\emailAdd{saikawa@hep.s.kanazawa-u.ac.jp}
\emailAdd{alexv@unizar.es}

\abstract{
The post-inflationary Peccei-Quinn (PQ) symmetry breaking scenario provides a unique opportunity to pinpoint the QCD axion dark matter mass, which is a crucial input for laboratory experiments that are designed for probing specific mass ranges. Predicting their mass requires a precise knowledge of how axions are produced from the decay of topological defects in the early Universe that are inevitably formed.

In this contribution, we present recent results on the analysis of the spectrum of axions radiated from global strings based on large scale numerical simulations of the cosmological evolution of the PQ field on a static lattice \cite{Saikawa:2024bta}.

We highlight several systematic effects that have been overlooked in previous works, such as the dependence on the initial conditions, contaminations due to oscillations in the spectrum, and discretisation effects; some of which could explain the discrepancy in the current literature.

Taking these uncertainties into account and performing the extrapolation to cosmologically relevant string tensions, we find that the dark matter mass is predicted to be in the range of 
$95\,\mu\text{eV} \lesssim m_a \lesssim 450 \, \mu\text{eV}$, which will be probed by some of the next generation direct detection experiments.
}

\FullConference{%
  2nd Training School and General Meeting of the COST Action COSMIC WISPers (CA21106)\\
  11-14 June and 3-6 September, 2024\\
  Ljubljana (Slovenia) and Istanbul (Turkey)}

\begin{document}
\maketitle

\section{Introduction}
\noindent
The quantum chromodynamics (QCD) axion\footnote{For a thorough discussion of the topic we refer to recent reviews on axion cosmology~\cite{OHare:2024nmr},  axion models~\cite{DiLuzio:2020wdo} and experimental searches~\cite{Irastorza:2018dyq}.}, originally proposed as a solution to the \emph{strong CP problem}~\cite{Peccei:1977hh,Peccei:1977ur,Weinberg:1977ma,Wilczek:1977pj}, is a leading candidate to explain cold dark matter (CDM) in our Universe~\cite{Davis:1986xc,Arias:2012az}.

In the post-inflationary Peccei-Quinn (PQ) symmetry breaking scenario, a network of global cosmic strings inevitably forms in the early Universe when the global continuous PQ symmetry is spontaneously broken, as explained by the Kibble mechanism~\cite{Kibble:1976sj}. These strings decay, radiating axions that contribute to the dark matter density today. Understanding the axion production from strings offers the unique opportunity to give a precise prediction of the axion dark matter mass, which serves as crucial input for experimental searches, as it determines the frequency ranges for resonant detection methods such as haloscopes. However, accurate predictions are hindered by uncertainties on the axion emission spectrum, particularly its spectral index \(q\), and the evolution of the string network density. Even with extensive numerical simulations, the system can only be studied at unphysical regimes that ultimately require careful extrapolations over many orders of magnitude in the string tension. Previous studies yielded conflicting results, with predictions for the axion mass ranging from a few to $\sim 1000\ \mu$eV \cite{Gorghetto:2018myk, Gorghetto:2020qws, Buschmann:2021sdq, Hindmarsh:2021zkt, Kim:2024wku, Benabou:2024msj}. 
In this contribution we discuss the relevant parameters that characterize the axion radiation from global strings and several systematic effects that could explain the discrepancy in the existing literature. 

\section{Simulating Axion Strings}
\noindent 
To study the dynamics of the network of global axion strings we simulate a complex scalar field $\phi$ with Lagrangian,
\begin{equation}
\mathcal{L}=\frac{1}{2}\left|\partial_\mu \phi\right|^2-V_{\mathrm{PQ}}(\phi), \quad\text{where}\quad V_{\mathrm{PQ}}(\phi)=\frac{\lambda}{4}\left(|\phi|^2-f_a^2\right)^2.
\end{equation}
Here, $f_a$ defines the high-energy scale of the spontaneous symmetry breaking. Expanding the field around the potential minima located at $|\phi|=f_a$ yields
$\phi(x)=\left(f_a+r(x)\right) e^{i \theta(x)}$.
We identify $a(x)= \theta(x)f_a$ as the \emph{axion}, i.\,e. the massless Goldstone boson, and $r(x)$ as the heavy radial field, dubbed \emph{saxion}, with mass $m_r = \sqrt{2\lambda}f_a$.\\
Strings are located where the modulus \( |\phi| \) approaches zero, as the phase \( \theta(x) \) winds through a full circle and continuity of the field requires \( |\phi| = 0 \) at one point within each loop in $\theta$, where \( \theta \) is therefore undefined. These regions have high energy densities as they correspond to maxima of the potential in field space, i.\,e. they retain residual potential energy from the unbroken phase $\langle\phi\rangle=0$.

To run simulations of the PQ field during radiation domination in an expanding FRW background, we utilise \texttt{jaxions}\footnote{Available on Github: \url{https://github.com/veintemillas/jaxions}}\cite{jaxions}, a massively parallel \texttt{C++} code, originally developed in Ref.~\cite{Vaquero:2018tib}. Performing numerical simulations of global axion strings suffers from an infeasible issue with the dynamical range of the system. This becomes most apparant when looking at the \emph{energy per unit length} (or \emph{tension}), $\mu$, of the string, which is logarithmically divergent:
\begin{equation}
   \mu = \hat{\mu}+\pi f_a^2\int_{c_{\mathrm{UV}}}^{c_\mathrm{IR}}\frac{\mathrm{d}r}{r}\approx\hat{\mu}+\pi f_a^2\ln\left(\frac{c_\mathrm{IR}}{c_{\mathrm{UV}}}\right). \label{eq:string_tension}
\end{equation}
In the cosmological scenario of the QCD axion, the UV cutoff $c_{\mathrm{UV}}$ is provided by the width of the string core $c_{\mathrm{UV}}\approx m_r^{-1}$ and the IR cutoff scale by the typical inter-string distance $c_\mathrm{IR} \sim H^{-1}$, which is motivated by the fact that the strings are expected to enter a \textit{scaling} regime with $\mathcal{O}(\text{few})$ strings per Hubble patch \cite{Kibble:1976sj,Yamaguchi:1998iv}.
At the latest times $f_a/H\sim 10^{30}$, and therefore $\ln\left(\frac{f_a}{H}\right)\sim 70$, which is \emph{impossible} to resolve, even on the largest supercomputers. This poses a major problem as we are forced to simulate at unphysical, low values of the string tension and extrapolate results over several orders of magnitude.

From this point we use the short-hand notation $\ell \equiv \ln\left(\frac{f_a}{H}\right)$ for the parameter that we use to track the temporal evolution of the network, since $\ell\equiv\ell(\tau)$ grows with conformal time, because $H\propto 1/\tau^2$ during radiation domination.

\section{String Network Density and Attractor}
\begin{figure}[t]
    \centering
    \raisebox{-0.5\height}{\includegraphics[width=0.49\textwidth]{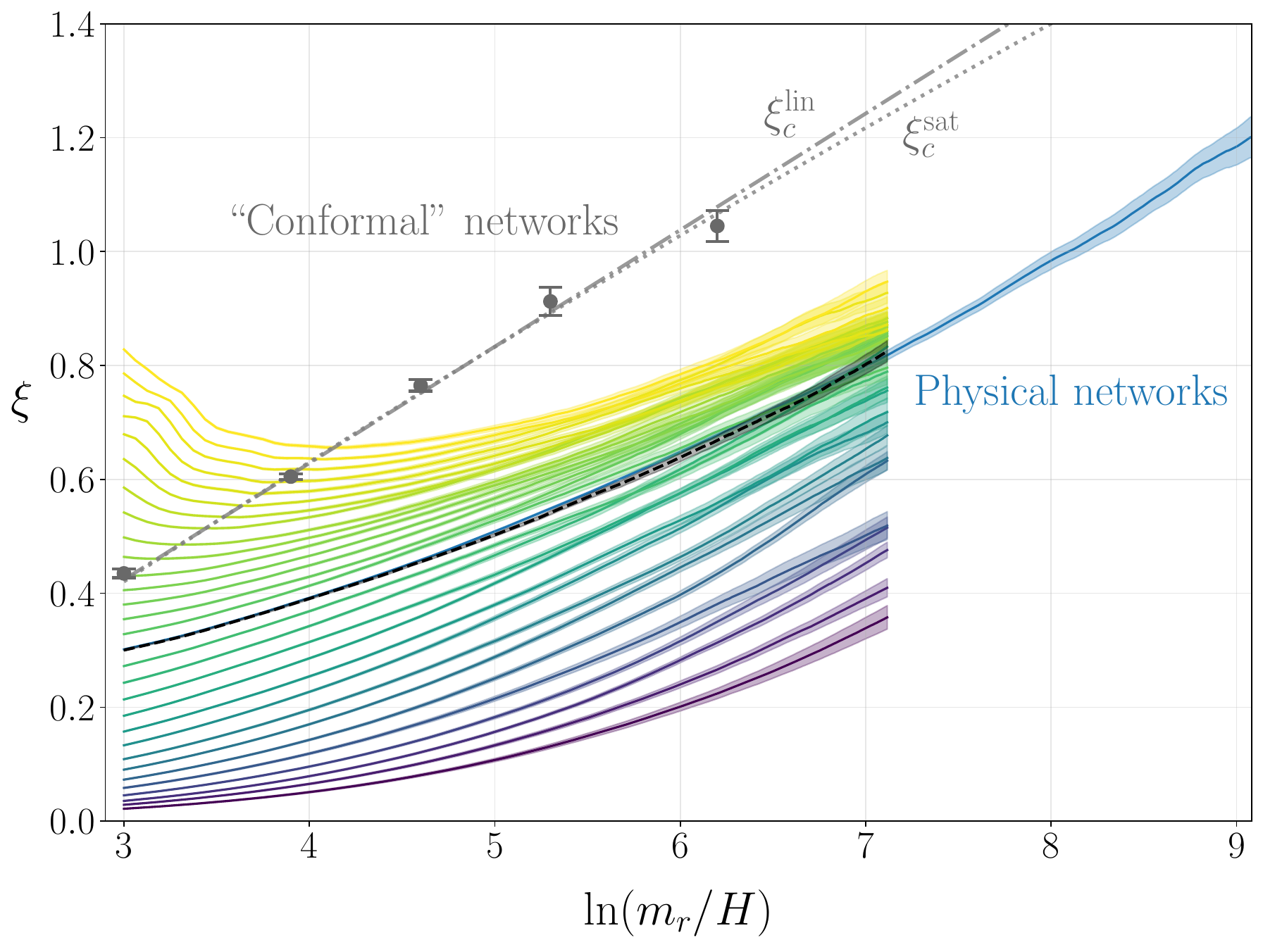}}
    \hfill
    \raisebox{-0.5\height}{\includegraphics[width=0.48\textwidth]{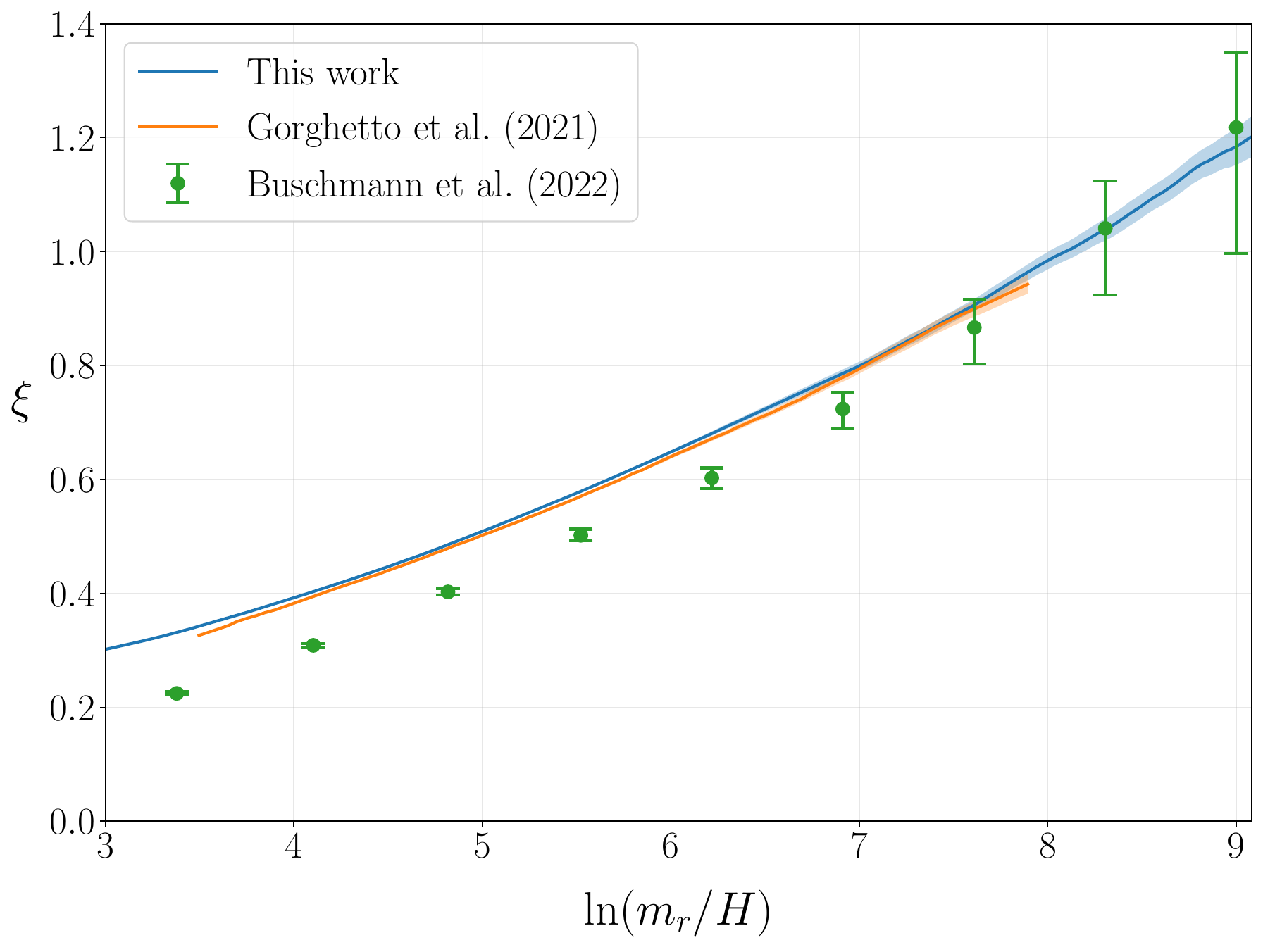}}
    \caption{\textit{Left:} Evolution of the string density $\xi$ for different initial densities (simulations with $N^3=2048^3$ grid sites) and for the largest simulations ($N^3=11268^3$) using initial conditions close to the attractor $\xi\approx 0.3$ (black dashed line).  \textit{Right:} Comparison of the string density evolution with the results of Refs. \cite{Gorghetto:2020qws,Buschmann:2021sdq}. Figures from Ref. \cite{Saikawa:2024bta}.}
    \label{fig:xi}
\end{figure}

\noindent
Determining the evolution of the string density parameter $\xi = \frac{l_s}{\mathcal{V}}t^2$, with $l_s$ the string length and $\mathcal{V}=t^3$ the causal volume in conformal units, is very important to minimize the impact of the choice of initial conditions on the axion dark matter mass prediction. Previous studies, cf. Ref. ~\cite{Gorghetto:2018myk}, observed and introduced the notion of an \emph{attractor} solution of the string density evolution. We confirmed the existence of such behavior, cf. Fig. \ref{fig:xi}, and found that it can also be observed in the energy spectra of both the axion and especially the saxion fields.

The attractor evolution, inspired by the study of conformal string networks\footnote{Conformal string networks are artificially kept at fixed values of $\ell$.} of Ref. \cite{Klaer:2019fxc}, can be modeled as, 
\begin{equation}
\frac{d \xi}{d t}=\frac{C}{t}\left(\xi_c(\ell(t))-\xi(t)\right).
\label{eqn:xi_model}
\end{equation}
Here, $\xi_c(\ell)$, is the equilibrium density of the conformal string network and $t/C$ is the characteristic time scale of the network restoration, approaching the limit $\xi_c\rightarrow\xi$.

In modifying the constant $C\rightarrow C(x=\xi/\xi_c)$, two models give satisfactory fits to the conformal data of Ref. \cite{Klaer:2019fxc}. Applying this to our simulations yields 
\begin{equation}
 \xi_c^{\mathrm{lin}}=-0.19(3)+0.205(7) \ell\quad \text{and}\quad 
 \xi_c^{\mathrm{sat}}=\frac{-0.25(15)+0.23(6) \ell}{1+0.02(4) \ell}.
\end{equation}
Evaluated at $\ell=70$, this results in values of
\begin{equation}
\xi^{\operatorname{lin}}(\ell=70)  \sim 13.8(5)\quad \text{and}\quad \xi^{\mathrm{sat}}(\ell=70)  \sim 7(3),
\end{equation}
disfavouring an early saturation as proclaimend in Refs. \cite{Hindmarsh:2019csc, Hindmarsh:2021vih, Hindmarsh:2021zkt}, that find $\xi\sim 1.2(2)$. A comparison with the results of Refs. \cite{Gorghetto:2020qws, Buschmann:2021sdq} is displayed in the right panel of Fig. \ref{fig:xi}.

Identifying the attractor is crucial for the extrapolation. We found that for simulations of networks with lower initial string densities, there are more contributions from small scales rendering the radiation spectrum more IR-dominated, whereas overdense networks give rise to more UV-dominated spectra. The characteristics of the axion spectrum will be discussed subsequently.

\newpage
\section{Axion Spectrum}
\begin{figure}[t]
    \centering
    \raisebox{-0.5\height}{\includegraphics[width=0.49\textwidth]{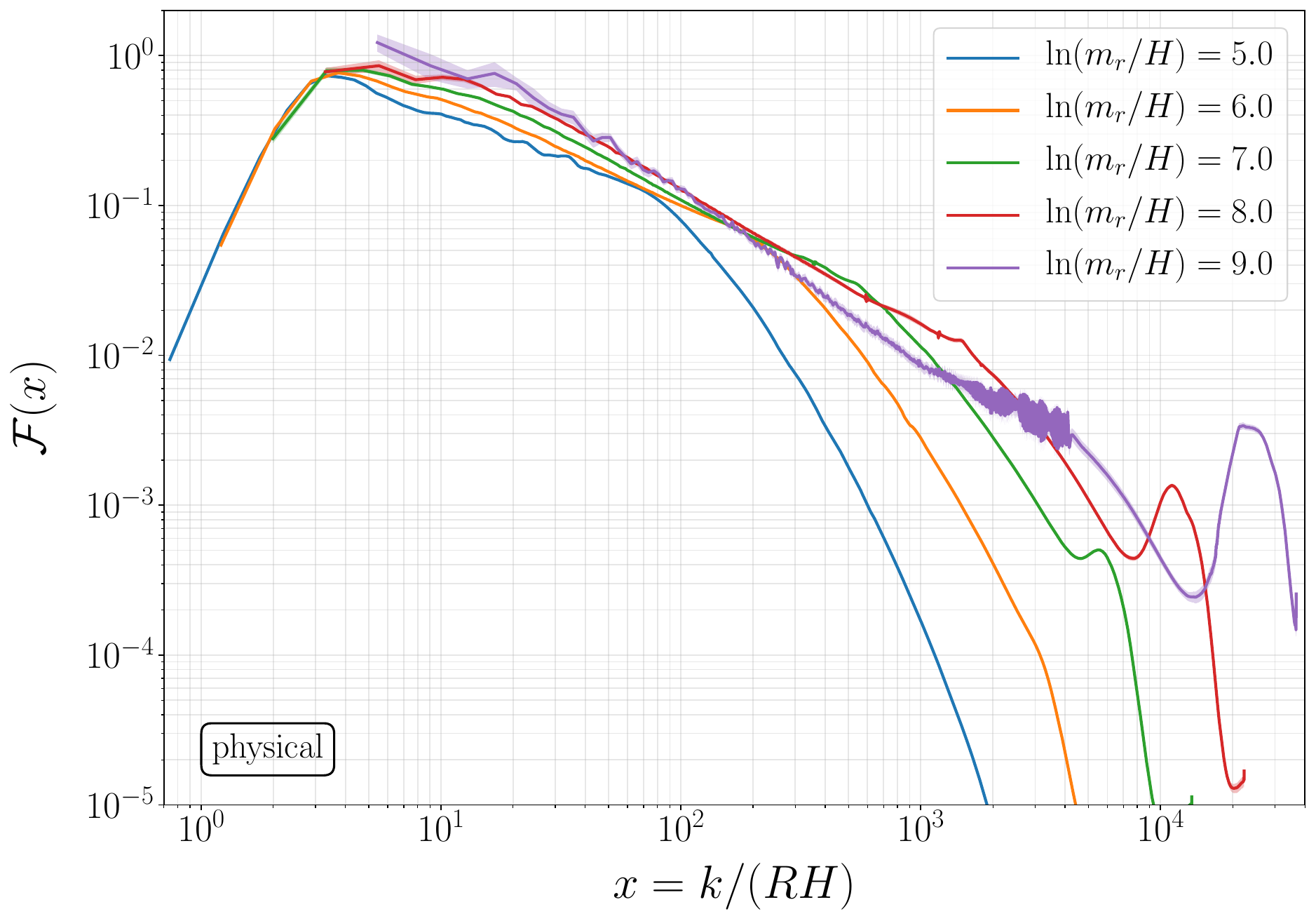}}
    \hspace{0.02\textwidth}
    \raisebox{-0.5\height}{\includegraphics[width=0.47\textwidth]{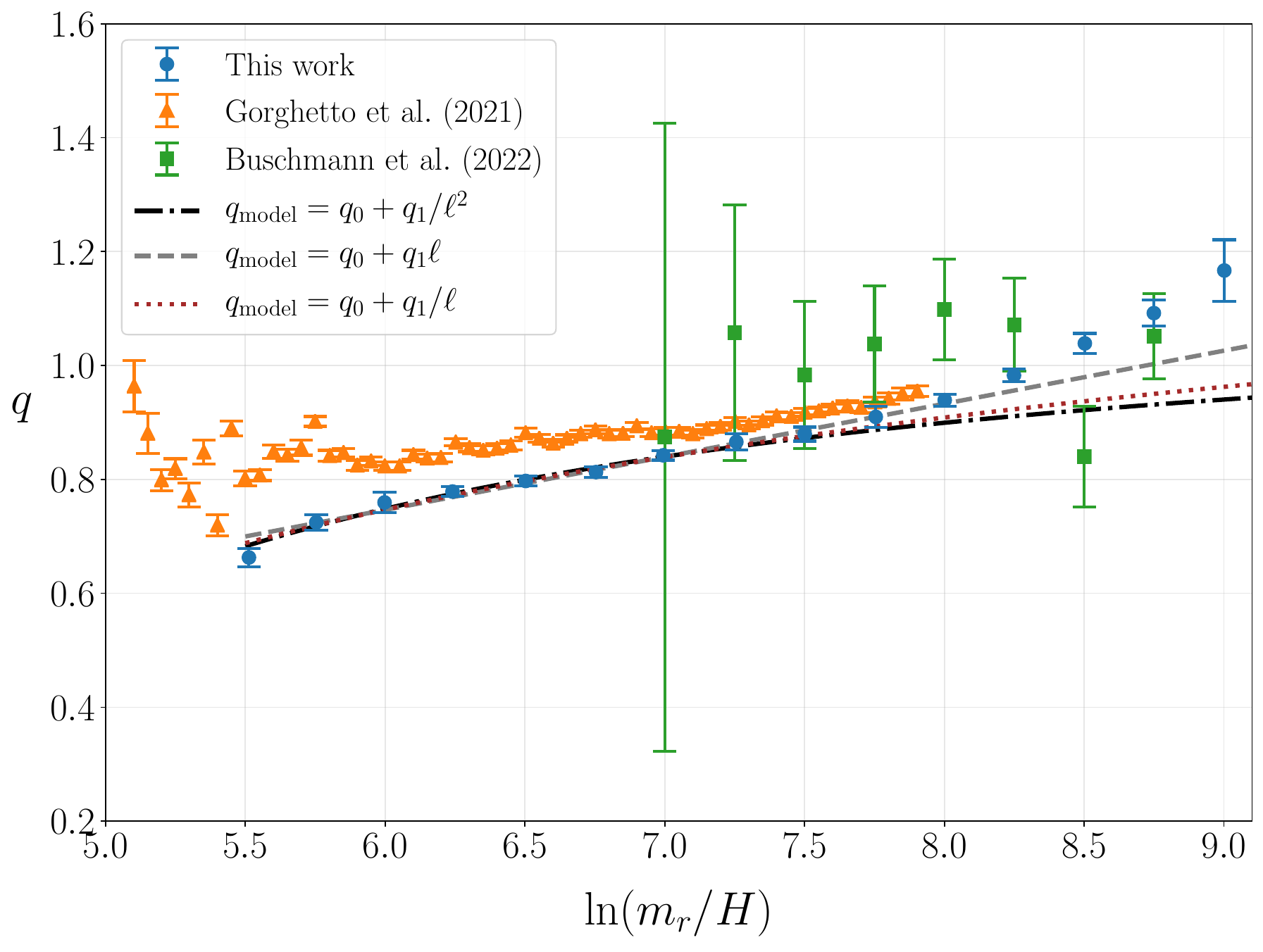}}
    \caption{\textit{Left:} Instantaneous axion emission spectrum $\mathcal{F}(x)$ for different times during the evolution. The increased distortion of the spectrum due to discretization effects is clearly visible for the last spectrum (purple). \textit{Right:} Evolution of the spectral index $q$ with fits compared to the results of Refs. \cite{Gorghetto:2020qws,Buschmann:2021sdq}. Figures from Ref. \cite{Saikawa:2024bta}.}
    \label{fig:spectrum}
\end{figure}

\noindent In modern terminology, the axion radiation from strings is characterized by the dimensionless \textit{instantaneous emission spectrum},   
\begin{equation}
\mathcal{F}(x, y)  \equiv \frac{1}{\left(f_a H\right)^2} \frac{\partial \Gamma_a}{\partial(k / R)} =\frac{1}{\left(f_a H\right)^2} \frac{1}{R^3} \frac{\partial}{\partial t}\left(R^4 \frac{\partial \rho_a}{\partial k}\right),
\end{equation}
where \(k\) is the wavenumber and \(R\) is the scale factor. We use the shorthand notation $x=k/(RH)$ and $y=m_r/H$ and further
\begin{equation}
\frac{\partial \Gamma_a(k)}{\partial k}=\left(f_a H\right)^2 \frac{k^3}{2 \pi^2} \frac{d \mathcal{N}(k)}{d t},
\end{equation}
as the \emph{spectral production rate}, with $\mathcal{N}(\boldsymbol{k})=\left(\left|\partial_\tau \widetilde{\psi}(\boldsymbol{k})\right|^2+k^2|\widetilde{\psi}(\boldsymbol{k})|^2\right) /(2 k \mathcal{V})$, the angle-averaged conserved occupation number for the axion plane wave solutions of the form $\psi(\boldsymbol{x})=\int\frac{d^3\boldsymbol{k}}{(2\pi)^3}\widetilde{\psi}(\boldsymbol{k})e^{i\boldsymbol{k}\cdot\boldsymbol{x}}$. The energy density in massless axion waves is therefore given by,
\begin{equation}
\rho_a=\left(f_a H\right)^2 \int d k \frac{k^3}{2 \pi^2} \mathcal{N}(k).
\end{equation}
The spectrum is expected to follow a power law behaviour of the form, 
\begin{equation}
    \mathcal{F}=\begin{cases}\mathcal{F}_0x^{-q}&(x_0<x<y),\\0&(\mathrm{otherwise}),
\end{cases}
\end{equation}
characterised by the \textit{spectral index} $q$.
This general shape can be observed in the left panel of Fig. \ref{fig:spectrum}. 
Knowing the exact value of $q$ is very important, as it determines if $\mathcal{F}$ is IR-dominated ($q>1$, i.e. many soft axions and therefore increased DM abundance) \cite{Gorghetto:2018myk, Gorghetto:2020qws, Kim:2024wku} , UV-dominated ($q<1$, fewer hard axions, suppressed DM abundance) or even a conformal spectrum for the intermediate case $q=1$, as strongly preferred by the recent results of Refs. \cite{Buschmann:2021sdq, Benabou:2024msj}.
There are several systematic effects that, besides the choice of initial conditions discussed in the previous section, could potentially bias the value of $q$:
\begin{enumerate}
\item \textbf{Oscillations in the spectrum}:
The evolution of the system naturally results in the emergence of a \emph{coherently} oscillating axion field when the relevant mode crosses the horizon or when the radial field triggers a parametric resonance effect. These oscillations can introduce contamination into the instantaneous emission spectrum and cause systematic errors if not properly accounted for during analysis.

\item \textbf{Discretisation effects}:
The limited resolution of the string core, caused by the finite lattice spacing, can distort the axion emission spectrum. This effect is conveniently characterized by the parameter $m_r a$~\cite{Fleury:2015aca}, defined as the product of the mass $m_r$ of the radial component of the PQ field (the inverse of the string core radius) and the lattice spacing, $a \equiv \frac{L}{N}$. Due to computational constraints, simulations near the continuum limit $m_r a \to 0$ are not feasible. As a result, larger values of $m_r a$ are typically used, which tend to overestimate the value of $q$. This problem is even more severe towards the end of the simulations as $m_ra$ increases to critical values of $m_ra\gtrsim 1.0$ due to the fact that the strings effectively shrink in the comoving simulation volume\footnote{To overcome this issue some simulations use the ``PRS" trick \cite{Press:1989yh}, which consists of keeping the value of $m_ra$ artificially fixed throughout the simulation. }. Additionally, a poor resolution of the discretized Laplacian, characterized by the number of neighbouring lattice points $N_g$ taken into account, tends to bias the value of $q$ towards higher values.
\end{enumerate}
Taking all of this into account, the evolution of the spectral index can be modeled following the general ansatz:
\begin{equation}
    \large q=q_{\mathrm{model}}(\ell)+q_{\mathrm{disc}}(\ell,m_ra),
\end{equation}
where $ q_{\mathrm{disc}}(\ell,m_ra)=d_0(m_ra)^{d_1}\exp\left[\frac{d_2\ell}{1+(m_ra)^{d_3}}\right]$, with some constants $d_i$ determined by the fit, which captures the effects discussed above and a total of four models (A-D) for the late time evolution of the spectral index give reasonable fits to the simulation data, namely:
\begin{equation}
\begin{aligned} 
\text{A:}\ q_{\mathrm{model}}&=q_0+q_1\ell\quad(q_1>0), &&\text{B:}\ q_{\mathrm{model}}=q_0+q_1\ell^2\quad(q_1>0),\\ \text{C:}\ q_{\mathrm{model}}&=q_0+q_1/\ell\quad(q_1<0), &&\text{D:}\ q_{\mathrm{model}}=q_0+q_1/\ell^2\quad(q_1<0).
\end{aligned}
\end{equation}
In the right panel of Fig. \ref{fig:spectrum} we display the results for the different fit models compared to the data of Refs. \cite{Gorghetto:2020qws, Buschmann:2021sdq}. 
\section{Axion Relic Abundance and Dark Matter Mass}
\noindent
The number density of axions, and therefore their relic abundance, is directly related to the shape of the spectrum via 
\begin{equation}
 \frac{n_{a}^{\mathrm{str}}}{f_{a}^{2}H} =\frac{1}{f_a^2H}\int dk\frac{1}{\omega}\frac{\partial\rho_a}{\partial k} =\int^\tau\frac{d\tau^{\prime}}{\tau}\int\frac{dx^{\prime}}{x^{\prime}}\mathcal{F}(x^{\prime},y^{\prime}),
\end{equation}
where $\omega=k/R$.
\begin{figure}[t]
    \centering
    \includegraphics[width=0.49\textwidth]{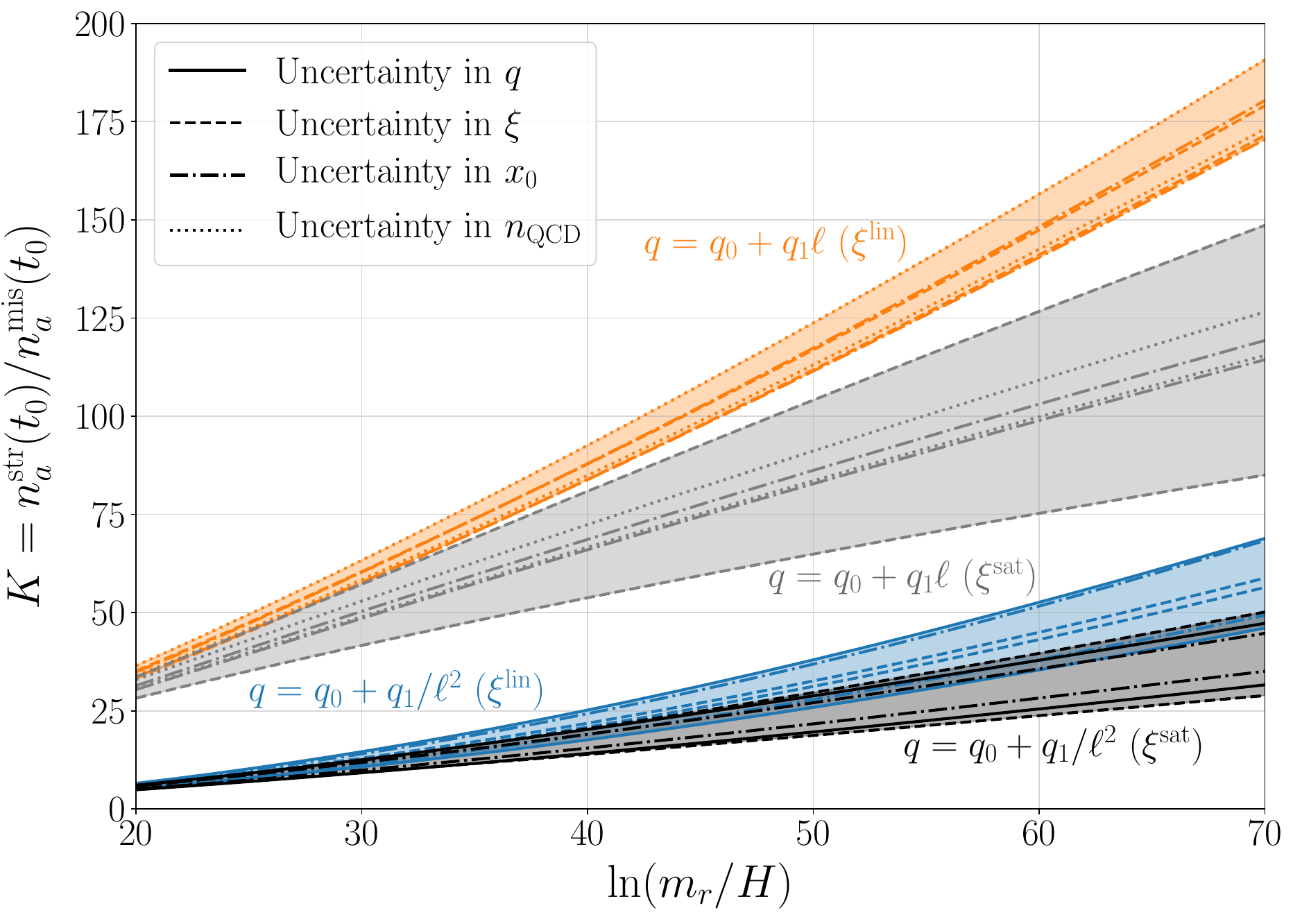}
    \includegraphics[width=0.49\textwidth]{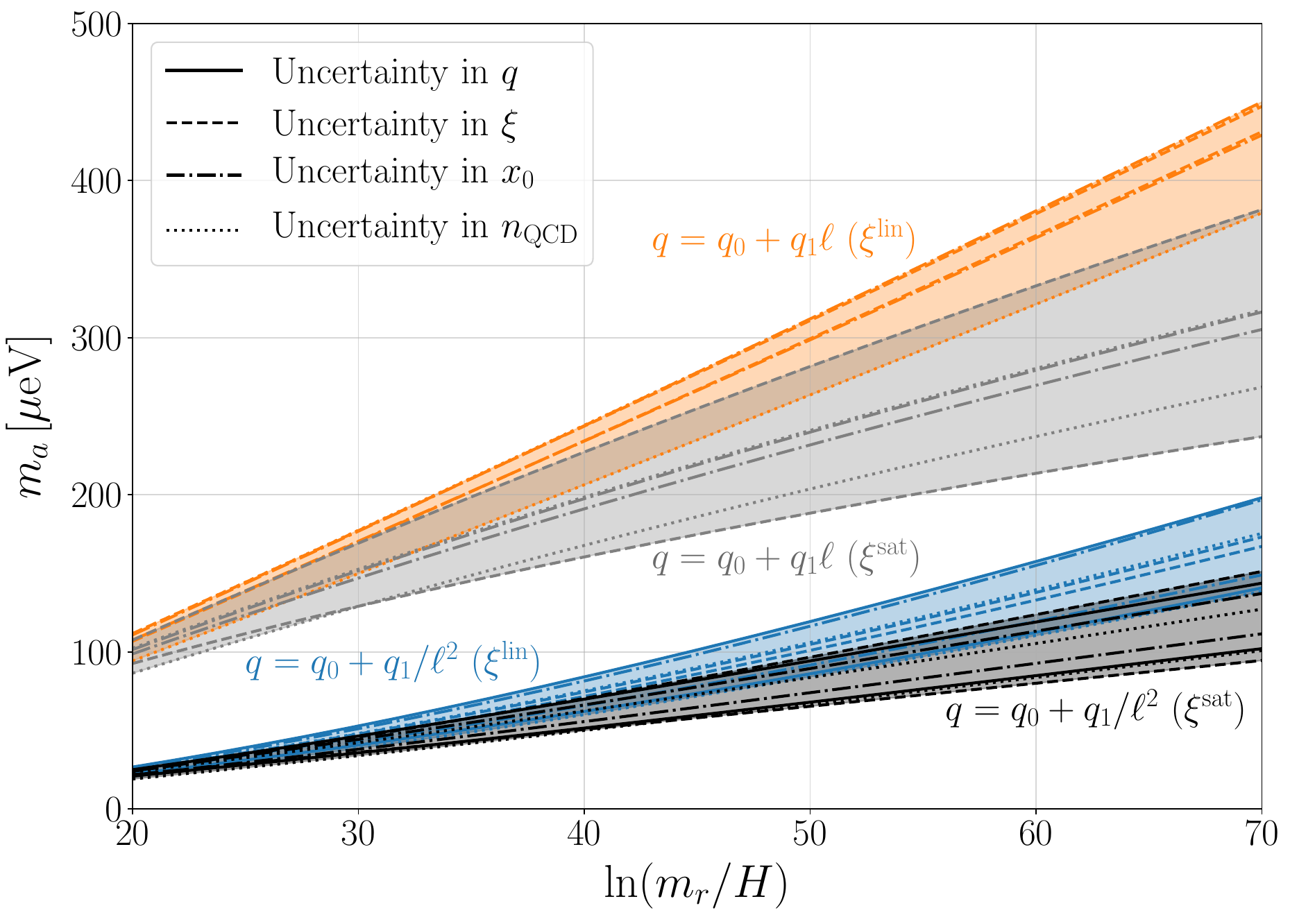}
    \caption{\textit{Left:} Result for the different extrapolations of the numerical parameter $K$, that quantifies the axion production from topological defects over the standard misalignement production. \textit{Right:} Extrapolation of the axion dark matter mass for the different models. Figures from Ref. \cite{Saikawa:2024bta}.}
    \label{fig:extrapol}
\end{figure}
In order to quantify the effect of strings on the estimation of the dark matter abundance,
let us define the production efficiency from strings compared to the usual misalignment estimate as
\begin{align}
K \equiv \frac{n_a^{\rm str}(t_0)}{n_a^{\rm mis}(t_0)}, \label{K_definition}
\end{align}
where $n_a^{\rm mis}(t_0) = c_{\rm mis}H_1f_a^2 \left(R_1/R_0\right)^3$ and $c_{\rm mis} = 2.31$
is the result for the angle-average misalignment estimate~\cite{Borsanyi:2016ksw,GrillidiCortona:2015jxo}.
In terms of $K$, the axion CDM abundance can be estimated as
$\Omega_a h^2 = K\Omega_a^{\rm mis}h^2$, where
$\Omega_a^{\rm mis}h^2 = \frac{m_a n_a^{\rm mis}(t_0)}{\rho_{\rm crit}/h^2}$.
For a given value of $K$, one can then determine the axion mass $m_a$ satisfying the condition $\Omega_ah^2 = \Omega_{\rm CDM}h^2 = 0.12$~\cite{Planck:2018vyg}.

Taking into account the different extrapolation schemes discussed earlier, the results for the parameter $K$ can be found in the left panel of Fig. \ref{fig:extrapol}. These results can directly be casted into the prediction of the axion dark matter mass, which are displayed in the right panel of Fig. \ref{fig:extrapol}. 

\section{Conclusions}
\begin{figure}[t]
    \centering
    \includegraphics[width=\textwidth, trim=90 0 0 800, clip]{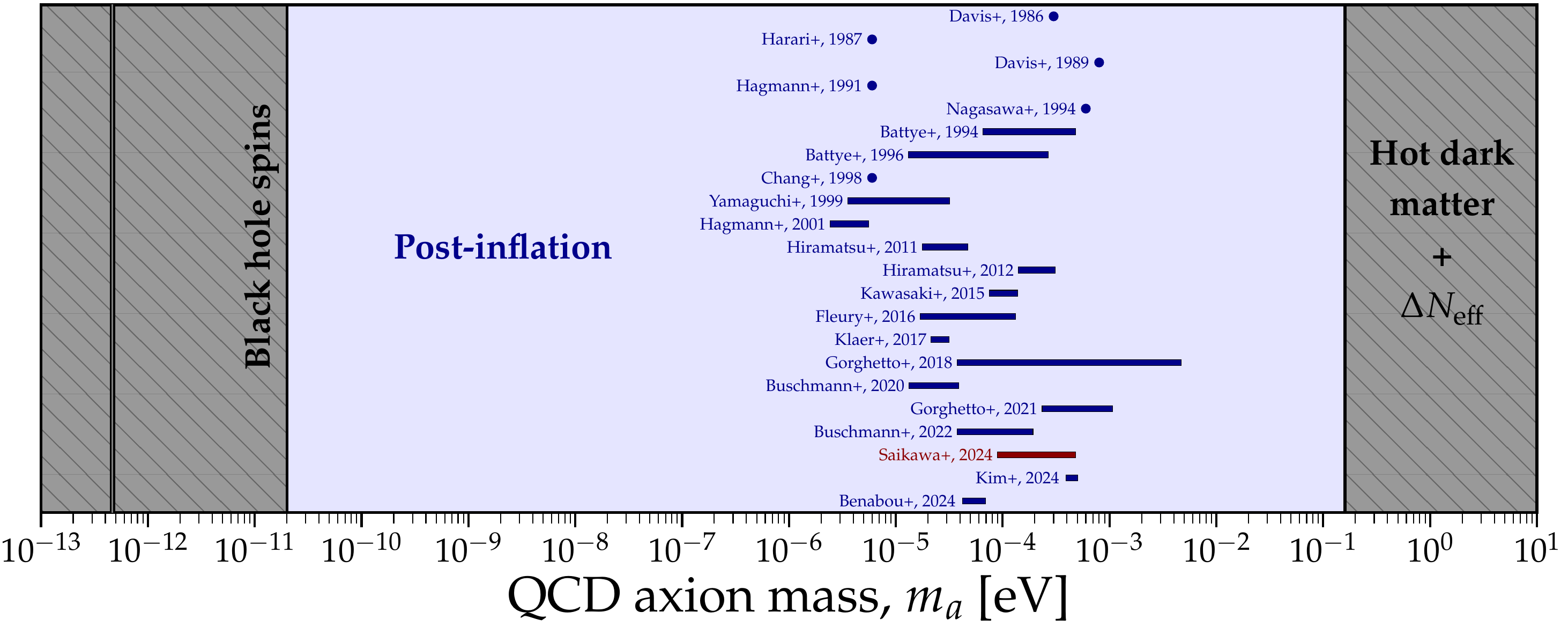}
    \caption{Overview of predictions for the axion dark matter mass from string simulations in the post-inflationary scenario. The figure is adapted from Ref. \cite{Saikawa:2024bta} (result highlighted in red) and updated with the latest results from Refs. \cite{Kim:2024wku, Benabou:2024msj}.  The original template is courtesy of Ciaran O'Hare \cite{AxionLimits}.}
    \label{fig:mass_predictions}
    \hfill
\end{figure}
\noindent We discussed recent results on the dynamics of global string networks using large scale simulations of the axion dark matter field.  

The existance of an  attractor behaviour for the string density parameter $\xi$ was confirmed and modeled by a simple differential equation, cf. Eqn. \ref{eqn:xi_model}, minimising potential bias on the determination of the spectral index $q$ from initial conditions. 

The instantaneous emission spectrum is largely contaminated by oscillations in the axion energy density from misalignment oscillations of the Fourier mode of the axion field after the corresponding mode enters the horizon and   parametric resonance effects induced by the oscillating radial field. Careful numerical treatment allows for a reduction of the impact of these oscillations in the computation of the instanteneous emission spectrum $\mathcal{F}(x)$.

Most importantly, we quantified discretisation effects on the spectrum, which bias $q$ towards larger values. These effects become more severe towards the end of the simulations, at larger values of $\ell=$ $\ln \left(m_r / H\right)$. 

Taking all the above considerations into account, multiple global fits of the relevant parameters are performed and the axion dark matter mass is predicted in the range of 
\begin{equation}
    95\,\mu\text{eV} \lesssim m_a \lesssim 450 \, \mu\text{eV}.
\end{equation}
An overview of how this prediction compares to previous studies can be found in Fig. \ref{fig:mass_predictions}. 

The predicted mass range aligns well with the sensitivity of upcoming experiments, including  FLASH \cite{Alesini:2023qed}, (Baby)IAXO \cite{Armengaud:2014gea}, RADES \cite{Melcon:2018dba}, ADMX \cite{ADMX:2020ote} and QUAX \cite{QUAX:2020adt} at the lower end of the predicted mass (or frequency $\sim m_a/2\pi$) range and, even if experimentally more challenging, for example ALPHA \cite{ALPHA:2022rxj}, MADMAX \cite{Caldwell:2016dcw} or ORGAN \cite{McAllister:2017lkb} at the upper end. These results provide a theoretical foundation for optimizing experimental search strategies, focusing efforts on the \(\mu\text{eV}\) mass range.

A possible next step to increase the dynamical range of the simulations is to use adaptive mesh refinement (AMR), as successfully demonstrated in Refs. \cite{Buschmann:2021sdq, Benabou:2024msj}, for string networks and, in Refs. \cite{Drew:2019mzc, Drew:2022iqz, Drew:2023ptp}, for individual string loops. Alongside the progress with the numerical setup, there is also ongoing work to obtain a deeper theoretical understanding of the radiation associated with different global string loops \cite{Kaltschmidt2025}.

\section*{Acknowledgements}
\noindent 
This article is based on work from COSMIC WISPers (CA21106), 
supported by COST (European Cooperation in Science and Technology). 
Part of this research was supported by the Munich Institute for Astro- and Particle Physics (MIAPP) which is funded by DFG
under Germany's Excellence Strategy -- EXC-2094 -- 390783311.
M.\,K., J.\,R. and A.\,V. are supported by the grants PGC2022-126078NB-C21 funded by MCIN/AEI/ 10.13039/501100011033 and “ERDF A way of making Europe” and the DGA-FSE grant 2020-E21-17R Aragon Government and the European
Union - NextGenerationEU Recovery and Resilience Program on `Astrof\'{i}sica y F\'{i}sica de Altas Energ\'{i}as CEFCA-CAPA-ITAINNOVA. Additionally, the work of M.\,K. is supported by the Government of Aragón, Spain, with a PhD fellowship as specified in ORDEN CUS/702/2022. The work of K.S. is supported by JSPS KAKENHI Grant Number JP24K07015. A.V. is further supported by AEI (Spain) under Grant No. RYC2020-030244-I / AEI / 10.13039/501100011033.
Computations were performed on the Raven and Cobra HPC systems at the Max Planck Computing and Data Facility.
This work was achieved in part through the use of SQUID at the Osaka University Cybermedia Center.

\bibliographystyle{utphys}
\providecommand{\href}[2]{#2}\begingroup\raggedright\endgroup

\end{document}